# Unsupervised Classification of Uncertain Data Objects in Spatial Databases Using Computational Geometry and Indexing Techniques

## Kurada Ramachandra Rao*, PSE Purnima*, M Naga Sulochana*, B Durga Sri**

*(Department. of Computer Applications, Shri Vishnu Engineering College for Women, Bhimavaram
** (Department of Computer Science and Engineering, Sri Sai Aditya Institute of Science and Technology, Surampalem

**ABSTRACT**

**Unsupervised classification called clustering is a process of organizing objects into groups whose members are similar in some way. Clustering of uncertain data objects is a challenge in spatial data bases. In this paper we use Probability Density Functions (PDF) to represent these uncertain data objects, and apply Uncertain K-Means algorithm to generate the clusters. This clustering algorithm uses the Expected Distance (ED) to compute the distance between objects and cluster representatives. To further improve the performance of UK-Means we propose a novel technique called Voronoi Diagrams from Computational Geometry to prune the number of computations of ED. This technique works efficiently but results pruning overheads. In order to reduce these in pruning overhead we introduce R\*-tree indexing over these uncertain data objects, so that it reduces the computational cost and pruning overheads. Our novel approach of integrating UK-Means with voronoi diagrams and R\* Tree applied over uncertain data objects generates imposing outcome when compared with the accessible methods.**

## 1. Introduction

Unsupervised classification [11] is a clustering technique where no predefined classes exist. Clustering [11] is a collection of data objects. Grouping is a set of data objects into cluster is called cluster analysis. A superior clustering method produces high quality cluster with high intra-class similarity and lower inter-class similarity. Clustering applications are widely used in pattern recognition, spatial data analysis, Document classification, and Real world application like Marketing, City planning etc.

The primary goal of unsupervised classification is to minimize the sum of squared error by minimizing the distance between the data object and the cluster representative. Clustering when applied to a mobile node distributed sensor networks and wireless technology [5] forms a group network with a cluster representative and cluster members. The cluster representative exchanges data and centroid information with the server etc. [14] in the form of batch mode for efficient telecommunication.

Short ranged signal has higher bandwidth and is used for local communications in with the clusters. Long ranged communication with the cluster representative and members also needs high bandwidth which is feasible by batch transmission [5], [14]. Mobile nodes construct and report their locality by comparing the strength of radio signal within the mobile access point inside the large ranged communication, that may sometimes create noise, and such devices in mobile computing practically are called uncertain, whose locations are updated in the database sporadically. In certain instance of time, the location is not known the latest update value is considered as the sampling time instance for uncertain data [2], [4], [17] by considering various geometrical constraints.

Each uncertain data attribute is a type, subject to its own independent probability distribution called attribute uncertainty. In correlated uncertainty, multiple attributes are described by joint probability distribution. In tuple uncertainty, all the attributes of a tuple are subjected to this probability distribution [4].

In this paper we consider the unsupervised learning of uncertain objects where locations are uncertain and, hence they are defined by Probability Density Functions (PDF). Since the conventional clustering methods works with point valued data, the uncertain data object has to be transferred into the same point valued data so that any conventional algorithm can be applied [13]. To handle clustering data objects we consider the object PDF rather than the conventional methods since they give better clustering.

We assume each object lies within the region and is bounded by finite bounding box. The PDF is zero outside the region. The uncertain data objects are first iterated through K-Means algorithm in a iterative procedure. The Euclidean distance is used to find the closeness between the cluster members and cluster representatives. The same uncertainty data objects are experimented using Uncertain K-Means [13], [15] instead of computing the Euclidean distance the Expected distance (ED) is used to compute the centroid between the cluster representative and cluster members. Expected distance involves the numerical integration using large number of sample points for each PDF, so that the computational cost is reduced.

In this paper we also introduce one of the computational geometry called voronoi diagram [16], [8], [9] so that it can prune some of the candidate clusters. This pruning technique is used to consider the spatial relationship [12] among the cluster representatives. We also prove voronoi diagrams based pruning is far efficient than the boundary box techniques. For efficiency in ED computing, we also apply pruning based on boundary box based technique over the objects to establish lower and upper boundary for ED. The proposed voronoi diagram pruning technique prunes the ED and thus saves the computational cost and impacts the execution time.

Spatial indexing methods are used to process magnanimous spatial database [11], [12] for fast and effective results. These indexing methods directly effect the memory efficiency of spatial data [7] as well as the spatial retrieval performance. To reduce the disk space for spatial uncertain data we use R*-tree proposed by Gutman [10], which adopts the smallest bounding rectangular (MBR) to divide spatial entity by using "the smallest area criterion", and construct dynamic index tree. R*- tree [1], [18] organizes the spatial index according to the data, at the same time, it is the balanced tree using MBR to express the objects, and the nature expansion highly based on B-tree in k-dimension.

Each node in R*-tree contains the uncertain object, which are represented as rectangular regions in space MBR. Here the Voronoi based pruning techniques are applied to the entire rectangular region visited on a single uncertain object. This shows the combination R*-tree and voronoi techniques significantly reduces the pruning overheads.

This paper is organized as follows. Section 2 reviews the Related work, the problem and the proposed solutions in Section 3. The detailed experimental setup and results are shown in Section 4, while Section V concludes the work and gives directions to future work.

## 2. Related Work

In recent years, uncertain data has become ubiquitous [5], it is often associated with uncertainty because of inaccurate measurement inaccuracy, sampling discrepancy, outdated data sources, or other errors. For example, in the scenario of moving objects (such as vehicles or people), it is impossible for the database to track the exact locations of all objects at all-time instants. Therefore, the location of each object is associated with uncertainty between updates [14].

These various sources of uncertainty have to be considered in order to produce accurate query and mining results. We note that with uncertainty, data values are no longer atomic. To apply traditional data mining techniques, uncertain data has to be summarized into atomic values. Taking moving-object applications as an example again, the location of an object can be summarized either by its last recorded location or by an expected location. Unfortunately, discrepancy in the summarized recorded value and the actual values could seriously affect the quality of the mining results. In recent years, there is significant research interest in data uncertainty management.

Data uncertainty [2], [4] can be categorized into two types, namely existential uncertainty and value uncertainty. In the first type it is uncertain whether the object or data tuple exists or not. For example, a tuple in a spatial database could be associated with a probability value that indicates the confidence of its presence. In value uncertainty, a data item is modeled as a closed region which bounds its possible values, together with a probability density function of its value. This model can be used to quantify the imprecision of location and sensor data in a constantly-evolving environment. In this paper we study the problem of clustering objects with tuple uncertainty [3], [15].

UK-means algorithm a generalization of K-means algorithm is to handle objects where locations are uncertain. The location of each object is described by the probability density function (PDF). The UK-means also computes the expected distance between each object and the cluster representative. For arbitrary PDF, calculation the ED between the object and a cluster representative is represented as an integration computation. We consider various pruning methods to avoid such expensive ED calculations. One of the pruning techniques proposed was Min-Max Bounding Box (MM-BB) distance pruning technique that reduces the computational cost.

The other technique uses the voronoi diagrams [8]. The voronoi diagram is one of the most fundamental and versatile data structure in computational geometry. The voronoi diagram divides a space into disjoint polygons where the nearest neighbor of any point inside a polygon is the generator of the polygon. The role of voronoi diagrams in the context of clustering is many folds. For certain applications, the relevant cluster structure among the objects is well reflected, in a direct manner, by the structure of the Voronoi diagram of the corresponding point sites [6]. For instance, dense subsets of sites give rise to Voronoi regions of small area (or volume). Regions of sites in a homogenous cluster will have similar shape. For clusters having a direction-sensitive density, the regions will exhibit an extreme width in the corresponding direction. Perhaps more important is the fact that numerous types of optimal clustering are induced by Voronoi diagrams. The clustering minimizing the sum of the squared distances of the clusters to their centers is easily found by constructing the voronoi diagram.

The R*-tree [1] the variant of R-tree is a state-of-the-art spatial index structure. It has already found its way into commercial systems like SQLite, MySql and Oracle. The most important improvement of the R*-tree over the original R-tree is that it utilizes forced reinsertion. That is, if a disk page overflows, some objects are removed from the page and reinserted into the index. The R* tree uses the same algorithm of R-tree for both query and delete operations. The goals are: to reduce the MBR area and to keep the shape of the MBR close to a square. The R*-tree algorithm selects objects whose distances to the center of the page's MBR are the largest. R*-tree groups the underlying points hierarchy and records the MBR of each group for answering spatial queries. In this paper we use R*-tree for indexing and improved split heuristic procedures [18].

## 3. PROPOSED WORK

A set of objects $O = \{o_1, \ldots, o_n\}$ is a m-dimensional space Rm with a distance function $d: R^m \times R^m \to R$ giving the distance $d(x,y) \geq 0$ between any points $x, y \in R^m$. is considered, associated with each object is a pdf $f_i: R^m \to R$, which gives the probability density of oiat each point $x \in R^m$. By definition of pdf, we have (for all i = 1,. . . ,n)

$f_i(x) \geq 0 \ \forall \ x \in R^m$ and $\int_{x \in R^m} f_i(x) \, dx = 1$

Further, we assume that the probability density of oi is confined in a finite region Ai, so that $f_i(x) = 0$ for all $x \in R^m \setminus A_i$. We define the expected distance between an object oi any point $y \in R^m$.

$$: ED(o_i, y) \int_{x \in A_i} d(x,y) \ f_i(x) \, dx \qquad (1)$$

Now, given an integer constant k, the problem of clustering uncertain data is to find a set of cluster representative points $C = \{c_1, ..., c_n\}$ $C = \{c1,..., cn\}$ and a mapping h: $\{1, ..., n\} \rightarrow \{1, ..., k\}$ so that the sum of squared expected distance is minimized.

$$\sum_{i=1}^{n} [ED(o_i, c_{h(i)})]^2$$

To facilitate our discussion on boundary box based algorithms, we use MBRi to denote the minimum bounding rectangle of object oi. MBRi is the smallest box, with faces perpendicular to the principal axes of Rm, which encloses Ai. Note that Equation (1) still holds if we replace "$x \in A_i$" with " $x \in MBR_i$". This fact can be overworked for optimization when computing ED.

First the UK-Means algorithm is applied to the spatial data sets. To reduce the computational cost and to do pruning, we apply the Min-Max Bounding Box technique to the UK-Means, later the proposed voronoi based pruning technique is implemented in to uk-means and finally to reduce pruning overheads we integrate R*-tree indexing algorithm in to the uk-means.

### 3.1 UK-Means
Clustering algorithms are based on k-means, in which the goal is to minimize sum of square error (SSE). The basic idea behind the uncertain k-means algorithm is to minimize the expected sum of squared errors. UK-Means algorithm is a generalized k-means algorithm to handle objects whose locations are uncertain. The UK-Means algorithm could be characterized as the least robust of all the methods, shown as Algorithm 1 UK-Means. Its insensitivity to variance within a distribution can be viewed as a major flaw, especially given that the distributions of the features in the cells are extremely variable.

| Algorithm 1 UK-Means |
|---|
| Step 1: Choose k arbitrary point as $c_j$ (j = 1,...,k) |
| Step 2: **repeat** |
| Step 3:     **for** all $o_i \in O$ **do**    // assign objects to cluster |
| Step 4:         **for** all $c_j \in C$ **do** |
| Step 5:             Compute ED($o_i, c_j$) |
| Step 6:             h(i) ← arg $\min_{j:c_j \in C} \{ED(o_i, c_j)\}$ |
| Step 7: **for** all j = 1,..., k **do** |
|                  // readjust cluster representatives |
| Step 8:         $c_j$ ← centroid of { $o_i \in O$ \|h(i) = j} |
| Step 9: **until** C and h become stable |

The basic drawback in this algorithm is that it computes ED for every object cluster pair in every cluster. So, given n objects, k clusters the UK-Means computes nk EDs in each iteration. The computation of an ED involves numerically integration a function that involve on object PDF. A PDF is represented probability distribution matrix, with each element representing a sample point in an MBR. To improve the performance of UK-Means, we need to reduce the time spent on ED. To avoid this ED we incorporate pruning into UK-Means.

We first apply the UK-Means algorithm to the spatial data sets [7]. To reduce the computational cost, we apply the MM-BB pruning technique in to the UK-Means, then we implement another effective pruning technique VCP to reduce pruning overheads we use indexing applied with R*-tree algorithm.

### 3.2 Min Max Bounding Box (MM-BB) Pruning
One of the pruning techniques we propose was to include MM-BB pruning in UK-Means, so that the computation cost of integration in ED will be reduced. In MM-BB MinMax pruning approach, for an object $o_i$ and a cluster representative cj, certain points in $MBR_i$ are geometrically determined. The distance from those points to $c_j$ are computed to establish bounds on ED. Formally, we define

$MinD(o_i, c_j) = \min_{x \in MBR_i} d(x, c_j)$
$MaxD(o_i, c_j) = \max_{x \in MBR_i} d(x, c_j)$
$MinMaxD(o_i) = \min_{c_j \in C} \{MaxD(o_i, c_j)\}$

It is apparent that $MinD(o_i, c_j) \leq ED(o_i, c_j) \leq MaxD(o_i, c_j)$. Then if $MinD(o_i, c_j) > MaxD(o_i, c_q)$ for some cluster representative $c_p$ and $c_q$, we can deduce that $ED(o_i, c_p) > ED(o_i, c_q)$ without computing the exact values of the EDs. So, object $o_i$ will not be assigned to cluster p since there is another cluster q that gives a smaller expected distance from object $o_i$. We can thus prune away cluster p without bothering to compute $ED(o_i, c_j)$. As an optimization, we can prune away cluster p if $MinD(o_i, c_p) > MaxD(o_i)$.

Now we include this MM-BB pruning technique shown as Algorithm 2 into the Algorithm 1by replacing the steps 5 and 6. The pruning condition $MinD(o_i, c_j) > MinMaxD(o_i)$ reduces many clusters, depending on data distribution. It avoids many ED computations over MinD and MaxD which are expensive.

We remarks that computing MinD and MaxD requires us to consider only a few points on the perimeter of an object's MBR, instead of all points in its pdf.

| **Algorithm 2** Min-Max Bounding Box (MM-BB) Pruning |
| --- |
| Step 1 : Choose k arbitrary point as $c_i$ (j=1,...,k) |
| Step 2 : **repeat** |
| Step 3 : **for** all $o_i \in O$ **do**   // assign objects to cluster |
| Step 4 : **for** all $c_j \in C$ **do** |
| Step 5 : Compute MinD($o_i,c_j$) and MaxD($o_i,c_j$) |
| Step 6 : Compute MinMaxD($o_i$) |
| Step 7 : **For** all $c_j \in C$ **do** |
| Step 8 : if MinD($o_i,c_j$) > MinMaxD($o_i$) then |
| Step 9 : Remove $c_j$ from $Q_i$ |
| Step 10: if $\|Q_i\|$ = 1 then    // only one candidate remains |
| Step 11: Compute ED($o_i,c_j$) |
| Step 12: h(i) ← $\arg\min_{j:c_j \in Q_i}$ {ED($o_i, c_j$)} |

### 3.3 Voronoi Cell Pruning (VCP)
The Min-Max based pruning significantly improves the performance of uk-means and efficiently evaluates the bound of ED and avoids many ED computations. The flaw in Min-Max boundary box was the technique does not consider any geometric structure of $R^m$ or spatial relationship among the cluster representative.

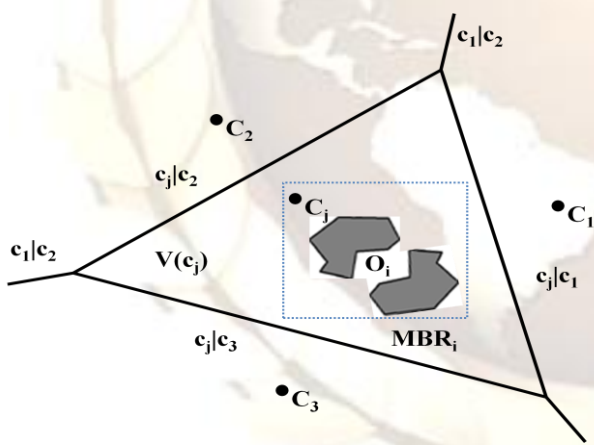

**Fig. 1 Voronoi-Cell Pruning (VCP)**

The Voronoi diagram is a fundamental structure in computational geometry and arises naturally in many applications including clustering. In this paper we use voronoi diagram [8] to build the spatial relationship between the cluster member and the cluster representative, and to achieve a very effective pruning. We compare the Min-Max bounding box pruning with voronoi pruning and we prove that voronoi diagram pruning is much stronger than the Min-Max bounding box technique.

Consider a set of points C={$c_1,...,c_k$}, the voronoi diagram divides the space $R^m$ into k cells $V(c_j)$ with the following property:

$: d(x, c_p) < d(x, c_q) \forall x \in V(c_p), c_q \neq c_p.$   (2)

The boundary of a cell $V(c_p)$ and its adjacent cell $V(c_q)$ consists of point on the perpendicular bisector, denoted $c_p|c_q$ between the points $c_p$ and $c_q$.

In all the iterations, we construct the voronoi diagrams from the k cluster representative points $C = \{c_1, ..., c_k\}$. The voronoi diagram is used to derive the VCP. For each object $o_i$, we check if $MBR_i$ lies completely inside any voronoi cell $V(c_q)$. If so, then object oi is assigned to cluster cj. This is because if follows from Equations (1) and (2) that:

$ED(o_i,c_j) < ED(o_i,c_q) \forall c_q \in C \{c_j\}.$

| **Algorithm 3**  Voronoi Cell Pruning (VCP) |
| --- |
| Step 1: Compute the Voronoi diagram for C={$c_1,...,c_k$} |
| Step 2: **For** all $c_j \in C$ **do** |
| Step 3:      **if**  $MBR_i \subseteq V(c_j)$  **then** |
| Step 4:           $Q_k \leftarrow \{c_j\}$ |

In this case, no ED is computed. All cluster except $c_j$ are pruned. An example for voronoi cell pruning is shown in Fig. 1, in which $V(c_j)$ is adjacent to $V(c_1), V(c_2), V(c_3)$. Since $MBR_i$ lies completely in $V(c_j)$, all points belonging to $o_i$ lie closer to $c_j$ than any other $c_q$. It follows that $ED(o_i,c_j)$ is strictly smaller than $ED(o_i,c_q)$ for all $c_q \neq c_j$. The pseudo code for VCP is shown in Algorithm 3. This code is embedded into Algorithm 1.

### 3.4 Indexing the uncertain objects
The two pruning technique proposed in this paper MM-BB, VCP aims to reduce the computational cost of ED, so that the execution time of uk-means improves. The results of algorithms are placed in section IV, where we observe about 75% of ED calculations is pruned and the computational cost is minimized by the pruning technique. Further to reduce the execution time pruning overheads we apply indexing over the uncertain objects.

Voronoi diagram based pruning technique takes the advantage of spatial distribution of cluster representative of the uncertain objects. The batch communications is used and nearby objects are grouped, we obtain MBR for each group. In order to save the computational time, groupings are done and arranged in a hierarchal order forming a super group and subgroups. The proposed technique uses top down approach to minimize the volume of MBR and we use R*-tree indexing for grouping the objects.

The R*-tree which is a variant of R-tree [1] is a self balancing tree like B+tree.  In both trees, the actual data either resides in the leaf nodes or is directly pointed to by the leaf nodes. The purpose of the intermediate nodes is to hold keys that partition and refine the node domain as one travels from the root node to the leaf nodes. This is especially well suited for spatial data because the data representing n-dimensional objects is often quite large. If this data were stored throughout the tree, as it is in a normal B+-tree, the nodes would only be capable of holding a few records and hence, driving the height of the tree unnecessarily high and decreasing performance of several operations. Locating the data, or pointers directly to the data, in the leaf nodes allows one to store more intermediate node records in fewer nodes, making the resulting tree height considerably lower.

Node insertion, deletion and splitting in B+-trees and R-trees and its variations are similar in basic concept. In the R* tree storage utilization heuristic is used and the forced-reinsert technique has been developed to implement this heuristic. The underlying consideration is that higher storage utilization will generally reduce the query cost as the height of the tree will be kept low.

The R*-tree is a hierarchical data structure. Each node corresponds to the smallest d-dimensional rectangle that encloses its child nodes. The leaf nodes contain pointers to the actual data in the database. Note that rectangles corresponding to different nodes may overlap. This means that a spatial query may often require several nodes to be visited before ascertaining the presence or absence of a particular rectangle.

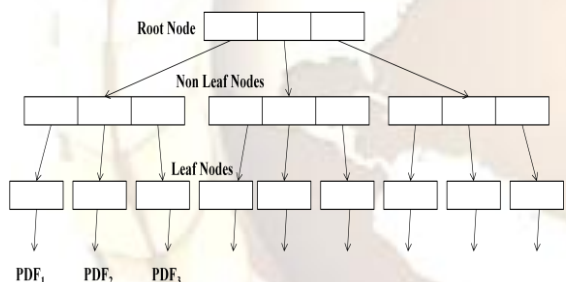

**Fig. 2 The file structure for the R*-tree with fan-out as 3**

In building an R*-tree, new rectangles are added to the leaf nodes. The appropriate leaf node is determined by traversing the R-tree starting at the root and at each step choosing the sub tree whose corresponding covering rectangle would have to be enlarged the least. Once the leaf node is determined, a check will be made to see whether the insertion will cause the node to overflow. If yes, then it must be split and the M+1records must be distributed in two nodes (where M is the order of the R*-tree). Splits are then propagated up the tree. Fig. 2 is a graphical representation of R*-tree used over the uncertain data objects with a fan out of 3 to each node.

Each tree node, containing multiple entries, is stored in a disk block. Based on the size of a disk block, the number of entries in a node is computed. The height of the tree is a function of the total number of objects being stored, as well as the fanout factors of the internal and leaf nodes. Each leaf node corresponds to a group of uncertain objects.

Each entry in the node maps to an uncertain object. The following information is stored in each entry:

- The MBR of the uncertain object.
- The centroid of the uncertain object.
- A pointer to the PDF data of the object.

The PDF data are stored outside the tree to facilitate memory utilization. Each internal node of the tree corresponds to a super-group, which is a group of groups. Each entry in an internal node points to a child group. Each entry contains the following information:

- The MBR of the child group.
- Number of objects under the sub tree at this child.
- The centroid of the objects under the sub tree at this child.
- A pointer to the node corresponding to the child

Storing the number of objects under the sub tree at a child node and the corresponding centroid location allows efficient readjustment of cluster representatives at the end of every iteration of UK-means. The R*-tree focuses its efforts on improving the accuracy of the data structure representation of spatial data by minimizing the following parameters when inserting and splitting nodes:

Area: This is the total area required to bind a set of objects minus the area covered by the objects. In other words, the area is the ``dead'' space in the bounding directory rectangles. Minimizing this produces a more compact tree which generally narrows the node domain that must be examined for each search operation.

Overlap: This is the area of intersection among data objects in the same node. Minimizing the overlap also minimizes the number of ``branches'' in the tree that must be visited for a search.

Margin: This is the sum of each bounding rectangle sides. Minimization of the margin value forces the splits toward producing more square bounding rectangles. More square bounding rectangles in turn pack better and improve the R*-tree quality and hence search operations.

The R*-tree also utilizes a forced reinsertion process in an attempt to alleviate the need for a node split and to improve the quality of the data organization. Forced reinsertion means that a set number of records in the full node are deleted from the node and reinserted in the tree. Reinsertion is invoked the first time a node overflows at the given tree level during the process of inserting the original record. This means at most, one node over flows at each level (excluding the root level) will be resolved using reinsertion, all other overflow in the reinsertion process will be handled by splits.

To build an R*-tree from a database of uncertain objects, we use a bulk-load algorithm based on the Sort-Tile-Recursive algorithm [10], [18]. It builds an R-tree from bottom up (as opposed to repeated insertion from the top and has the advantages of building a more fully filled tree, with smaller MBRs for the internal nodes, and a shorter construction time. Fig. 4 is the illustration how a MBR of 24 uncertain objects using the sort-tile recursive algorithms, with a fan out factor of leaf node as 3.

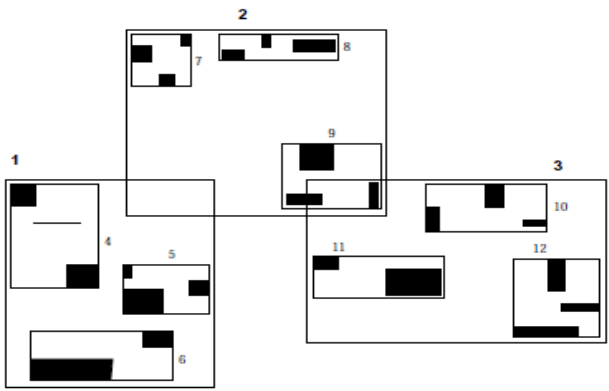

**Fig. 3  R*-tree with a Sort-Tile recursive procedure**

### 3.5 Group Based Pruning
Multilevel grouping of cluster objects is taken place in the R*-tree node. In order to increase the performance of pruning algorithm the pruning is applied in batch. We recursively traverse the tree from the root node to the leaf node by examining each entry in the node. Each entry e represents a group of uncertain objects. The MBR of e is available in R*-tree. Using this MBR we apply the pruning technique MM-BB, VCP and RMM-VCP to prune the cluster centers.

The pruning is done on the cluster representative $c_p$ if there is for sure another cluster representative $c_q \neq c_p$ such that all points in the MBR are closer to $c_q$ than to $c_p$. This property holds to all the subgroups and uncertain objects in the sub tree, which serves a lot of repeated computations and saves a lot of repeated computations.

In case only a single cluster representative $c_r$ is left then all descendants of e must be assigned to $c_r$. In this case, we can further optimize by bulk-assigning $c_r$ to the whole sub tree. There is no need to process each uncertain object under the sub tree individually. If this kind of sub tree pruning happens at higher levels of the R*-tree, a lot of processing can be saved. We now include this R*-tree algorithm in the Algorithm 1 with the recursive functions ProcessNonleafNode (r,c) and ProcessLeafNode(r,c), where r is the R*-tree's root node and c is the set of all clusters. The modified code in UK-means with Min-Max, Voronoi Cell and R*-tree functions give raise to the new pruning technique called RMM-VCP. One of the recursive procedure in RMM-VCP, ProcessNonleafNode(r,c) is shown as Algorithm 4.

The handling of leaf node is similar and hence not repeated. The procedure ProcessLeafNode differs from ProcessNonleafNode in which the recursive part (steps 7 - 11) is replaced by ED calculations and assigning the closest cluster to the uncertain object.

| **Algorithm 4** ProcessNonleafNode(r,c) |
|---|
| Inputs : n- R*-tree internal node, <br>           Q- a set of candidate clusters |
| Step 1 : **for** all child entry e of n **do** |
| Step 2 :     Apply pruning techniques MM-BB & VCP <br>              to Q using e's MBR |
| Step 3 : **if** \|Q\| =1 **then**   // only one cluster remains |
| Step 4 :     **for** all uncertain objects $o_i$ under subtree <br>                                    rooted at n **do** |
| Step 5 :         $h(i) \leftarrow j$ where $c_j \in Q$ |
| Step 6 : **else** |
| Step 7 :         $m \leftarrow e's$ R*-tree node |
| Step 8 : **if** m is leaf node **then** |
| Step 9 :         call ProcessLeafNode(m,Q) |
| Step 10: **else** |
| Step 11   call ProcessNonleafNode(m, Q) // recursively |

### 3.6 Hybrid Algorithms
In this paper we applied the pruning techniques MM-BB, VCP over the spatial databases and, for indexing in groups we used the R*-trees.  The UK-Means is embedded with the MM-BB pruning which raises the computational cost. We use a novel approach to combine with the VCP.  Candidate clusters are first pruned by the VCP.  If MM-BB is applied to an internal node N such that it reduces the set of candidate cluster representatives Q to a smaller set Q`, the reduced set Q` can be passed along to the child nodes of N where MM-BB is re-applied. This approach reduces the computational cost but raises pruning overheads. In order to reduce these overheads we apply the R*-tree indexing. The pruning achieved by MM-BB at different levels along a path of the R*-tree is thus acquisitive. The results are presented as tables and graphs for all the integrated techniques in Section IV.

### 4. Results
We used a PC with a CPU of Intel(R) core i3, 2.93GHz and 4GB RAM to implement the proposed algorithms using JDK1.6.0 on Windows 7 platform. For the computations of VCP we used the qhull programs. We considered the cluster shift operations in all the algorithms over the uncertain data objects in spatial databases.

4.1 Data Sets

We have used the spatial dataset from http://kdd.ics.uci.edu/databases/covertype/covertype.html. Forest CoverType is a benchmark problem in the UCI KDD Archive. This problem relates to the actual forest cover type for given observation that was determined from US Forest Service (USFS) Region to Resource Information System (RIS). Forest CoverType includes 581,012 samples represented in point valued data with 7 cover type, each sample has 54 attributes including 10 remotely sensed data and 44 cartographic data. We transform this data set into many uncertain data sets by replacing each data point with an MBR and also generate the PDF.  We experimented our algorithms with only 10 percent of the available data object as a training set.

**TABLE I : Parameters used in algorithms**

| Parameter | Description | Initial value |
|---|---|---|
| N | No. of uncertain objects | 20000 |
| K | No. of clusters | 50 |
| L | Max. side length of MBR | 2 |
| S | No. of samples per object | 128 |
| D | No. of dimensions | 2 |
| B | Block size of R*-tree node | 512 |

For each data set, a set of n MBRs are generated in m-dimensional space $[0,100]^m$. Each MBR's side length is generated randomly and is bounded by variable l. The MBR is divided into s grid cells, each corresponding to a PDF sample point. Each sample point is associated with randomly generated probability value, normalized so that the sum of probabilities of the MBR is equal to 1. These probabilities values give a discredited representation of the PDF $f_i$ of the corresponding object. For all the algorithms we use the same dataset with variable c as random points to serve as the initial cluster centers. The dataset and initial cluster representative will be fed as inputs to the algorithms. The parameters used for all the experiments are listed in Table I.

**4.2 Result of Algorithms**
We executed all the algorithms using the parameters listed in Table I over the spatial data set. The results are summarized in Table II.

**TABLE II : Results of algorithms over the Spatial Data Set**

| Algorithms (with cluster shift) | $t_I$ (ms) | $N_{ED}$ |
|---|---|---|
| MM-BB | 2861 | 0.74 |
| VCP | 2291 | 0.51 |
| RMM-VCP | 1010 | 0.62 |

The value $t_I$ is defined as the total execution time taken divided by the number of iterations executed. The value $N_{ED}$ is defined as the total number of ED calculations divided by the number of initialized objects by the number of iterations. It is evident from Table II that the proposed algorithm in the paper yields significant results. The VCP saves 20% of the execution time.

The RMM-VCP algorithm with an integrated approach of using R*-tree indexing with MM-BB and VCP saves the execution time by more than 50%. Pruning effectiveness of the algorithms can be examined by the smaller value in the $N_{ED}$ column, this is because in each iteration uk-means computes for each object all k expected distances from the object to the k cluster representatives.

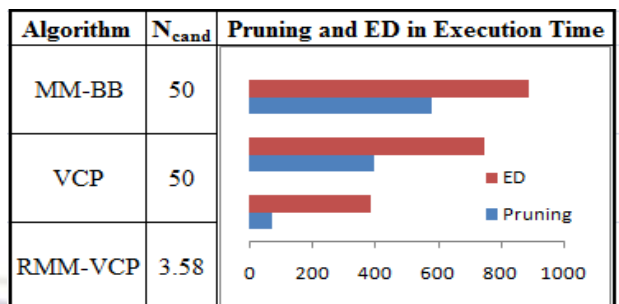

Fig. 4 Breakdown of ED and Pruning in Execution Time

An important observation is that the R*-tree does not affect the pruning effectiveness, but lowers the execution time. The execution time is involves both the time spent in ED calculations and pruning. The time spent in pruning involves a lot number of ED computations for MM-BB and checking against Voronoi cell boundaries for VCP. The numbers of such calculations are shown as $N_{cand}$, the average number of candidate object cluster pairs per iteration per object, which are shown in Fig 4.

**4.3 Effect of Number of Objects**
We used Fig. 5 to show effectiveness of the execution time in each iteration ($t_I$) over the uncertain objects (n). It can be seen that the execution time per iteration grows linearly with the number of uncertain objects. This is because as long as the pruning effectiveness and the effect of R*-tree boosting remains stable, the total number of ED computations and the pruning overheads will be proportional to the number of uncertain objects being handled.

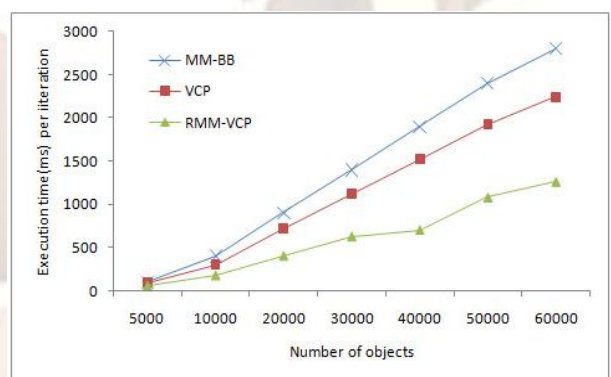

**Fig.5 Effect of No. of objects on Execution Time per Iteration**

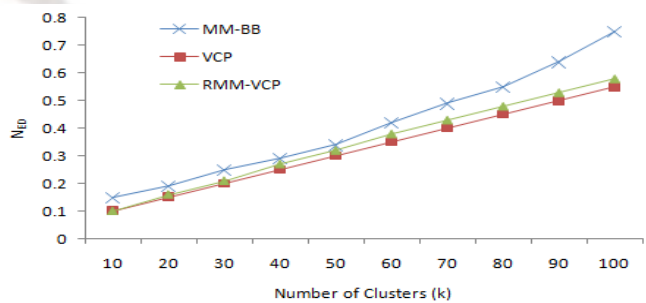

**Fig.6 Cluster formations using pruning techniques**

### 4.4 Effect of number of clusters
In Fig. 6 we consider the number of clusters k are varied from 1 to 100, from the graph we observe $N_{ED}$ increases with k. This is because in a larger number of clusters, cluster representatives are generally increased in number. Hence more ED will have to be computed to determine the cluster assignment. Fig. 6 shows that all the pruning techniques are very effective for a wide range of values of k in formation of clusters.

### 4.5 Effect of R*-tree Block size
Finally we test the effect of the block size of R*-tree nodes in the integrated algorithm. The block size effects the height of the R*-tree built, its compactness, the granularity of the groups and also the size of the MBR of each group. The results are shown in Fig. 7 and Fig. 8. The algorithms in Fig. 7 MM-BB and VCP do not employ R*-tree, and hence they do not effected by variations in block size of R*-tree. The other important observation was that execution time increases slightly with the block size in RMM-VCP algorithm.  From the Fig. 8 we notice that with smaller blocks, the number of nodes in the R*-tree increases, and so does the height of the R*-tree. This has a positive effect on pruning cost reduction because a deeper R*-tree allow more opportunities for batching the pruning computations, which can be applied to a larger number of nodes at more diverse granularities.

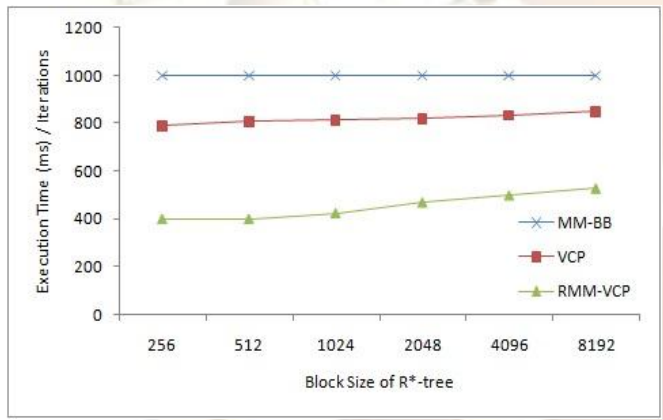

**Fig.7 Effect of Block size of R*-tree on Execution time per Iteration**

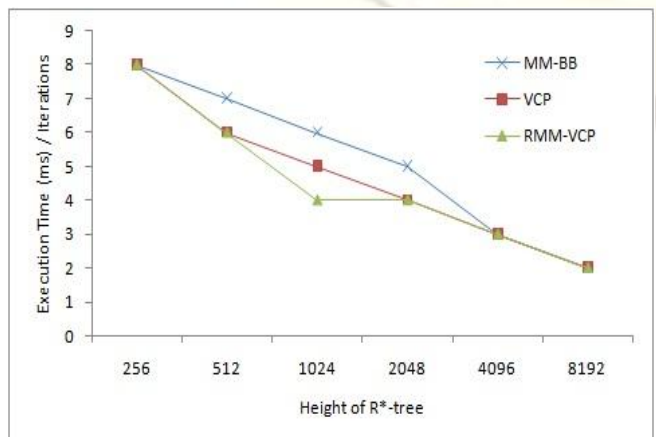

**Fig.8 Effect of Height of R*-tree on Execution time per Iteration**

### 5. Conclusion
In this paper we have analyzed about locations of uncertain objects by the PDF and clustering them. We used the UK-means algorithm with ED computations instead of other distance measures, since the number of samples used in representing the object PDF's are large. To improve the computational cost and apply pruning effectiveness on uncertain spatial data objects we used the MM-BB technique in UK-means. The drawback analyzed was they do not consider spatial relationship between the cluster members and cluster representatives. To further reduce the computational cost and improve pruning effectiveness we use the VCP. This algorithm gives a better result by reducing 97% of ED calculations, thus the execution time can be significantly reduced.

   For an optimal reduction in computational cost and perform spatial grouping, minimize pruning overheads on uncertain objects we used the R*-tree indexing which is a variant of R-tree. This indexing technique is incorporated with the MM-BB and VCP generates a new technique RMM-VCP for impressive pruning effectiveness. It is also proven in the previous section that this combination works well by outperforming the other approaches. Therefore we conclude that our innovative techniques based on computational geometry, indexing is reasonability competent.

   The future scope and enhancements and scope of this paper was to experiment the spatial data set with the density based clustering algorithms instead of partitioned based clustering algorithms, indexed with the other spatial data partitioning tree's like the x-tree, m-tree, Hilbert R-tree and Priority R-trees.